\documentclass[preprint,aps,amsmath,amssymb]{revtex4}

\newcommand {\bc}{\begin{center}}
\newcommand {\ec}{\end{center}}
\newcommand {\bea}{\begin{eqnarray}}
\newcommand {\eea}{\end{eqnarray}}
\newcommand {\be}{\begin{equation}}
\newcommand {\ee}{\end{equation}}

\def\lsim{\mathrel{\rlap{\lower4pt\hbox{\hskip1pt$\sim$}}
    \raise1pt\hbox{$<$}}}               
\def\gsim{\mathrel{\rlap{\lower4pt\hbox{\hskip1pt$\sim$}}
    \raise1pt\hbox{$>$}}}

\usepackage[mathscr]{euscript}
\usepackage{graphicx}

\begin{document}


\title{Viscosity spectral function of a scale invariant non-relativistic
fluid from holography}

\author{Thomas Sch\"afer}

\affiliation{Department of Physics, North Carolina State University,
Raleigh, NC 27695}

\begin{abstract}
We study the viscosity spectral function of a holographic 2+1
dimensional fluid with Schr{\"o}dinger symmetry. The model is based on 
a twisted compactification of $Ads_5\times S_5$. We numerically 
compute the spectral function of the stress tensor correlator for 
all frequencies, and analytically study the limits of high and low 
frequency. We compute the shear viscosity, the viscous relaxation 
time, and the quasi-normal mode spectrum in the shear channel. We 
find a number of unexpected results: The high frequency behavior 
is governed by a fractional 1/3 power law, the viscous relaxation 
time is negative, and the quasi-normal mode spectrum in the shear 
channel is not doubled. 
\end{abstract}

\maketitle

\section{Introduction}
\label{sec_intro}
 
 In recent years there has been a great deal of interest in the 
transport properties of strongly correlated fluids, such as the 
quark gluon plasma produced in collisions of relativistic heavy 
ions, the dilute Fermi gas at unitarity, and the strange metal phase 
of high $T_c$ superconductors \cite{Schafer:2009dj,McGreevy:2009xe,Adams:2012th,DeWolfe:2013cua,Schaefer:2014awa}. 
One of the central questions regarding these systems is whether the 
relevant degrees of freedom are quasi-particles, or whether the only 
possible description is in terms of non-local, holographic, degrees 
of freedom. This question is difficult to answer based solely on 
experimental data on the equation of state and the transport coefficients. 
Instead, it has been suggested that the problem can be addressed through 
the study of spectral functions, in particular the viscosity spectral 
function. If the excitations of the fluid are quasi-particles, then the 
spectral function has a quasi-particle peak at zero frequency. The
line shape is approximately Lorentzian, and the width of the quasi-particle
peak is related to the relaxation time. In holographic models, on the 
other hand, the spectral function is essentially featureless for
small frequency, and may even have a dip at $\omega=0$. There is no
direct relationship between the viscous relaxation time and the shape
of the spectral function. Poles of the retarded correlation function in
the complex plane are not related to quasi-particles, but to quasi-normal
modes. 

 In this work we consider the viscosity spectral function in a holographic
theory that describes a scale invariant non-relativistic fluid 
\cite{Son:2008ye,Balasubramanian:2008dm,Herzog:2008wg,Adams:2008wt,Maldacena:2008wh}. 
The theory is intended to serve as a model for the dilute Fermi gas 
at unitarity, which is a scale invariant strongly correlated fluid 
that can be studied using trapped ultracold atomic gases 
\cite{Bloch:2008zzb,Giorgini:2008zz}. Nearly perfect fluidity in the 
unitary gas was discovered in 2002 \cite{oHara:2002}, and a number of 
studies have demonstrated \cite{Schafer:2007pr,Cao:2010wa,Elliott:2013b} 
that near the phase transition between the normal and the superfluid phase 
the ratio $\eta/s$ of shear viscosity to entropy density of the unitary gas 
is close to the Kovtun-Son-Starinets bound $1/(4\pi)$ \cite{Kovtun:2004de}.

 For both high and very low temperatures the shear viscosity of the 
unitary Fermi is dominated by quasi-particles and can be computed using 
kinetic theory \cite{Bruun:2005,Rupak:2007vp}. Kinetic theory has also
been used to compute the viscosity spectral function \cite{Braby:2010tk},
and second order transport coefficients \cite{Schaefer:2014xma}. General
constraints on the spectral function are provided by sum rules 
\cite{Taylor:2010ju,Enss:2010qh}, and the high frequency behavior of 
the spectral function is governed by the operator product expansion
\cite{Hofmann:2011qs,Goldberger:2011hh}. Model independent constraints
at low frequency follow from hydrodynamics 
\cite{Chafin:2012eq,Romatschke:2012sf}, and lattice calculations of the 
viscosity spectral function have been reported in \cite{Wlazlowski:2013owa}.

 In the following we will study the spectral function of a non-relativistic
fluid in two spatial dimensions. Holographic models of 2+1 dimensional 
fluids can be constructed using light-like compactifications of $AdS_{5}
\times {\cal X}$, where ${\cal X}$ is a compact manifold
\cite{Son:2008ye,Herzog:2008wg,Adams:2008wt}. The basic idea can be 
explained based on the dispersion relation of a massless particle on 
the light cone, $p^+=p_\perp^2/(2p^-)$, where $\vec{p}_\perp=(p^x,p^y)$ is 
the transverse momentum and $p^\pm=p^0\pm p^z$ are light cone momenta. This 
dispersion relation exhibits Galilean scaling in 2+1 dimensions if the 
light-like momentum $p^-$ is discrete. 2+1 dimensional fluids have been 
studied experimentally, and hydrodynamic behavior was observed by studying 
the damping of collective modes \cite{Vogt:2011,Schafer:2011my,Bruun:2011}. 
We note that in 2+1 dimensions the scale invariance of a dilute classical 
Fermi gas is broken by a quantum anomaly \cite{Hofmann:2012np}, but the 
effect  of the anomaly on transport properties of the fluid is quite small 
\cite{Taylor:2012,Chafin:2013zca}.

 This paper is organized as follows. Th structure of the stress tensor
correlation function in fluid dynamics is reviewed in Sect.~\ref{sec_fluid}.
The Galilean invariant metric derived by Herzog et al.~\cite{Herzog:2008wg} 
and Adams et al.~\cite{Adams:2008wt} is reviewed in Sect.~\ref{sec_nrcft}, 
and the spectral function of the stress tensor correlation function is 
computed in Sect.~\ref{sec_eta_w}. We also study the low and high frequency 
limits, and compute the viscous relaxation time. In  Sect.~\ref{sec_qnm} we 
determine the spectrum of quasi-normal modes. Similar studies have been 
performed for the $AdS_5\times S_5$ black hole, which is dual to the ${\cal N}
=4$ supersymmetric Yang-Mills plasma. The viscosity spectral function was 
computed in \cite{Teaney:2006nc,Kovtun:2006pf}, and the viscous relaxation 
time was determined in \cite{Baier:2007ix}. The quasi-normal mode spectrum 
can be found in \cite{Starinets:2002br,Nunez:2003eq,Kovtun:2005ev}.

\section{Fluid dynamics}
\label{sec_fluid}
 
 The main focus of this study is the retarded correlation function
of the stress tensor, 
\be 
\label{G_R_xy}
 (G_R)_{xy,xy}(\omega,\vec{k})= -i \int dy\int d^2x\, 
 e^{i(\omega t-\vec{k}\cdot\vec{x})}\,  \Theta(t)\, 
 \left\langle \left[\Pi_{xy}(u,\vec{x}),\Pi_{xy}(0,0)\right]
 \right\rangle \, . 
\ee
The low energy, small momentum behavior of this correlation function 
is dictated by fluid dynamics. In fluid dynamics the stress tensor
of a scale invariant fluid is 
\be 
\label{Pi_ij_0}
 \Pi_{ij} = \rho u_i u_j + P g_{ij}+ \delta \Pi_{ij}\, ,
\ee
where $\rho$ is the mass density, $\vec{u}$ is the fluid velocity, $P$ 
is the pressure, $g_{ij}$ is the $d$-dimensional metric, and $\delta\Pi_{ij}$ 
contains terms that involve gradients of the thermodynamic variables. At 
first order in the gradient expansion $\delta\Pi_{ij}$ can be written as 
$\delta\Pi_{ij}=-\eta\sigma_{ij}$ with \cite{Son:2005tj} 
\be 
\label{sigma_ij}
 \sigma_{ij} = \nabla_i u_j +\nabla_j u_i + \dot{g}_{ij}
  -\frac{2}{d}g_{ij}   \langle\sigma\rangle \, ,
\hspace{0.1\hsize}
 \langle\sigma\rangle = \nabla\cdot u + \frac{\dot{g}}{2g}\, ,
\ee
where $\eta$ is the shear viscosity, $\nabla_i$ is a covariant derivative 
and $g=\det(g_{ij})$. At second order in the gradient expansion the stress
tensor is \cite{Chao:2011cy}
\bea 
\delta\Pi_{ij} &=& -\eta\sigma_{ij}
   + \eta\tau_\pi\left[
    g_{ik}\dot\sigma^{k}_{\; j} + u^k\nabla_k \sigma_{ij}
    + \frac{2}{d} \langle \sigma\rangle \sigma_{ij} \right] 
    + \lambda_1 \sigma_{\langle i}^{\;\;\; k}\sigma^{}_{j\rangle k} 
    + \lambda_2 \sigma_{\langle i}^{\;\;\; k}\Omega^{}_{j\rangle k}\nonumber\\
   && \mbox{} 
    + \lambda_3 \Omega_{\langle i}^{\;\;\; k}\Omega^{}_{j\rangle k}  
    + \gamma_1 \nabla_{\langle i}T\nabla_{j\rangle}T
    + \gamma_2 \nabla_{\langle i}P\nabla_{j\rangle}P
    + \gamma_3 \nabla_{\langle i}T\nabla_{j\rangle}P  \nonumber \\[0.1cm]
   \label{del_pi_fin}
   && \mbox{}
    + \gamma_4 \nabla_{\langle i}\nabla_{j\rangle}T 
    + \gamma_5 \nabla_{\langle i}\nabla_{j\rangle}P
    + \kappa_R  R_{\langle ij\rangle}\, . 
\eea
Here, ${\cal O}_{\langle ij\rangle}=\frac{1}{2}({\cal O}_{ij}+{\cal O}_{ji}
-\frac{2}{d}\delta_{ij}{\cal O}^k_{\;\;k})$ denotes the symmetric traceless 
part of a tensor ${\cal O}_{ij}$, $\Omega_{ij} = \nabla_iu_j-\nabla_ju_i$ 
is the vorticity tensor, and $R_{ij}$ is the Ricci tensor. 

 The low energy expansion of the retarded correlation is given by 
\cite{Chafin:2012eq}
\be 
\label{G_R_Kubo}
G_R(\omega,0)=P -i\eta\omega + \eta\tau_\pi \omega^2    + O(\omega^3) \, , 
\ee
where we have defined $G_R\equiv (G_R)_{xy,xy}$. Equation (\ref{G_R_Kubo})
can be used to determine the shear viscosity $\eta$ and the viscous 
relaxation time $\tau_\pi$. For non-zero momentum $k$ the shear mode
is diffusive and the dispersion relation is given by
\be 
\label{disp_shear}
 \omega = -i\nu k^2 -i\nu^2 \tau_\pi k^4 + \ldots \, ,
\ee
where $\nu=\eta/\rho$. Note that the $O(k^4)$ term is not complete, 
because it is suppressed by two powers of $k$ relative to the leading
order term, and at this level $O(\nabla^3)$ terms in the stress tensor
contribute. A popular scheme for implementing second order fluid 
dynamics, known as the Israel-Stewart method in the case of relativistic
fluids \cite{Israel:1979wp}, is based on promoting the viscous stress
tensor $\pi_{ij}\equiv\delta\Pi_{ij} $to a hydrodynamic variable. We can
write
\be
\label{pi_IS}
\pi_{ij} = -\eta\sigma_{ij}
   - \tau_\pi\left(
    g_{ik}\dot\pi^{k}_{\; j} + u^k\nabla_k \pi_{ij}
    + \frac{d+2}{d} \langle \sigma\rangle \pi_{ij} \right) 
    + \frac{\lambda_1}{\eta^2} \pi_{\langle i}^{\;\;\; k}\pi^{}_{j\rangle k} 
    + \ldots \, ,
\ee
where $\ldots$ refers to terms proportional to $\lambda_{2,3},\gamma_{1-3}$ 
and $\kappa_R$. Treating $\pi_{xy}$ as an independent variable, and 
solving the equations of linearized fluid dynamics we find one mode with 
the dispersion relation given in equ.~(\ref{disp_shear}), and a second 
mode described by 
\be 
\label{shear_IS}
 \omega=-\frac{i}{\tau_\pi}\, . 
\ee
Note that this mode is outside the low energy regime $\omega\ll\tau_\pi$. 
Finally, we can also study the dispersion relation of the sound mode. 
For this purpose we consider fluctuations of the form $\pi_{xx}(x,t)$.
We find 
\be 
\label{disp_sound}
 \omega = c_sk -\frac{i}{2} \Gamma k^2 +\frac{1}{8c_s} 
 \left( 8\left(1-\frac{1}{d}\right))c_s^2\nu\tau_\pi 
     -\Gamma^2\right) k^3 + \ldots \, , 
\ee
where the sound attenuation constant is 
\be 
\label{gamma_sound}
 \Gamma= 2\left(1-\frac{1}{d}\right) \nu
    + \frac{\kappa}{\rho} \left( \frac{1}{c_V}-\frac{1}{c_P}\right)\, .
\ee
Here, $\kappa$ is the thermal conductivity, and $c_{V,P}$ denotes the 
specific heat at constant volume and pressure, respectively. For simplicity, 
we have dropped $O(k^3)$ terms in equ.~(\ref{disp_sound}) that arise from 
$O(\nabla^2)$ terms in the entropy current \cite{Chao:2011cy}. In the 
Israel-Stewart scheme there is an additional mode with dispersion relation 
$\omega=-i/\tau_\pi$.

\section{Galilean invariant AdS/CFT}
\label{sec_nrcft}
 
 In order to study a holographic realization of non-relativistic fluid
dynamics we consider the metric constructed by Herzog et al.~and Adams 
et al.~\cite{Herzog:2008wg,Adams:2008wt}. We follow the notation of 
\cite{Herzog:2008wg}. The five dimensional metric is 
\bea 
\label{ds2_KK}
ds^2 &=& r^2k(r)^{-2/3} \left\{ 
   \left[ \frac{1-f(r)}{4\beta^2} - r^2f(r)\right] du^2
   + \frac{\beta^2r_+^4}{r^4}\, dv^2
   - \left[ 1+f(r)\right] \, du\, dv \right\} \nonumber \\
  & & \hspace{0.5cm}\mbox{} 
   + k(r)^{1/3} \left\{ r^2\, d\vec{x}^2 
      + \frac{dr^2}{r^2f(r)} \right\}  \, ,
\eea
where $u,v$ are light cone coordinates, $\vec{x}=(x_1,x_2)$, and 
$r$ is the radial $AdS$ coordinate. $\beta$ is a parameter that 
is determined the chemical potential, $r=r_+$ is the position of the 
horizon, and $f(r)=1-(r_+/r)^4$. We also define 
\be 
 k(r) = 1+\beta^2r^2\left( 1-f(r)\right) 
  = 1+ \frac{\beta^2r_+^4}{r^2}\, . 
\ee
This metric can be derived from the 10-dimensional metric describing 
non-extremal D3 branes via a null Melvin twist and Kaluza-Klein 
reduction. More straightforwardly, we can view equ.~(\ref{ds2_KK}) 
as a solution of the equations of motion for the five dimensional
action
\be 
\label{S_5d}
S = \frac{1}{16\pi G_5} \int d^5x\sqrt{-g} \left( 
  R-\frac{4}{3} (\partial_\mu\phi)(\partial^\mu\phi)
  - \frac{1}{4}e^{-8\phi/3} F_{\mu\nu}F^{\mu\nu} 
  - 4 A_\mu A^\mu - V(\phi) \right)\, ,
\ee
where the scalar potential is given by 
\be 
 V(\phi) = 4 e^{2\phi/3} \left( e^{2\phi} - 4 \right) \, . 
\ee
The classical solution for the the vector and scalar fields is given by
\be  
 A = \frac{r^2}{k(r)} \left( \frac{1+f(r)}{2} \, du 
    - \frac{\beta^2r_+^4}{r^4} \, dv \right)\, , 
 \hspace{0.5cm}
 e^\phi = \frac{1}{\sqrt{k(r)}}\, .
\ee
The Hawking-Bekenstein entropy of the black hole can be computed from 
the area of the event horizon, and the temperature follows from the 
surface gravity. We find 
\be 
S=\beta \frac{r_+^3}{4G_5}\,   \Delta v\Delta x_1\Delta x_2\, ,  \hspace{1cm}
T=\frac{r_+}{\pi\beta}\, . 
\ee
There is a non-zero chemical potential, $\mu=1/(2\beta^2)$, which is 
canonically conjugate to the momentum in the compactified $v$-direction. 
The equation of state is given $P\sim T^4/\mu^2$. This is an unusual 
equation of state, but consistent with scale invariance and stability. 
In particular, one finds ${\cal E} = P$, which follows from scale 
invariance in $2+1$ dimensions. In order for the density to be positive 
the chemical potential has to be negative, similar to a classical gas. 
The speed of sound is 
\be 
 c_s^2 =\left. \frac{\partial P}{\partial \rho}\right|_{s/n} = 
 \frac{|\mu|}{m}\, .
\ee
The speed of sound and the pressure go to zero in the limit $T\to 0$ at 
constant density. This means that in the zero temperature limit the 
equation of state of the holographic fluid behaves like a classical 
gas. In particular, there is no Fermi pressure and the Bertsch 
parameter \cite{Baker:1999np} is zero.

 The dilute Fermi gas has a phase transition to a superfluid state, which 
in the case of 2+1 dimensional gases is of Berezinskii-Kosterlitz-Thouless 
(BKT) type. The theory described by equ.~(\ref{ds2_KK}) does not exhibit
a phase transition, but it can serve as a model for the normal phase 
of a trapped atomic gas. The equation of state determines the density 
profile of a finite system confined by an external potential $V(x)\simeq 
\frac{1}{2} m\omega^2x^2$. In particular, the equation of hydrostatic
equilibrium, $\vec{\nabla}P=-n\vec{\nabla}V$ where $n$ is the density,
is solved by the local density approximation $\mu(x)=\mu_0-V(x)$. For
$P\sim T^4/\mu^2$ the density of a trapped gas is $n(x)\sim 1/[|\mu_0|
+V(x)]^3$, which is a physically reasonable model for a trapped gas.

\section{Viscosity spectral function}
\label{sec_eta_w}

\begin{figure}[t]
\bc\includegraphics[width=9cm]{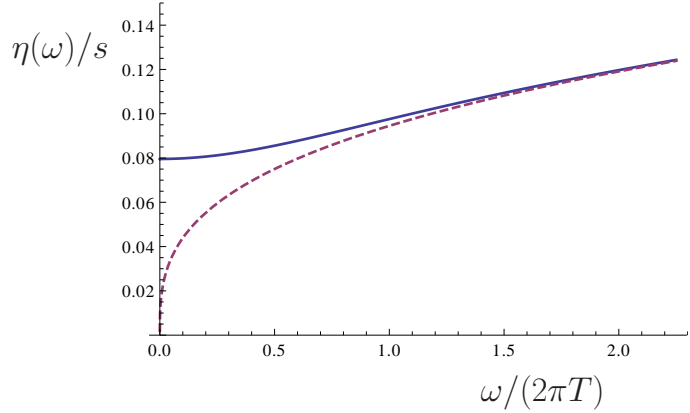}\ec
\caption{\label{fig_eta_w}
Viscosity spectral function $\eta(\omega)/s$ in the holographic
model constructed in \cite{Herzog:2008wg,Adams:2008wt}. The dashed
curve shows the asymptotic behavior $\eta(\omega)\sim \omega^{1/3}$. }   
\end{figure}

 We consider the retarded correlation function of the shear component
of the stress tensor
\be 
 (G_R)_{xy,xy}(\omega) = -i \int du\int d^2x\, e^{i\omega u}\, \Theta(u)\, 
 \left\langle \left[\Pi_{xy}(u,\vec{x}),\Pi_{xy}(0,0)\right]
 \right\rangle \, . 
\ee
In the holographic model this correlation function can be computed 
by studying fluctuations of the bulk metric of the form $\delta g_x^y
=e^{-i\omega u+inv} \chi(\omega,r)$. One can show that $n$ controls the 
particle number carried by the operator that is being probed. The 
stress tensor carries zero particle number and $n=0$. Small fluctuations 
are governed by the wave equation
\be 
\label{scal_wv_eq}
\frac{f(r)}{r^5}\frac{d}{dr}\left( r^5 f(r) \frac{d\chi}{dr}\right) 
  + V_{\it eff}(r) \chi(\omega,r) \, ,
\ee
where $V_{\it eff}=f(r)^2 g^{uu}/g^{rr}=(1-f(r))\beta^2\omega^2/r^4$. 
Using the explicit form of $f(r)$ and defining $\mathfrak{w}=\omega/
(2\pi T)$ we can write 
\be 
\label{ads_nr_de}
 \chi^{\prime\prime}(\omega,u_R) 
   - \frac{1+u_R^2}{f(u_R)u_R}\chi^\prime(\omega,u_R)
   + \frac{u_R}{f(u_R)^2}\mathfrak{w}^2\chi(\omega,u_R) = 0 \, , 
\ee
where $u_R=(r_+/r)^2$ and $f(u_R)=1-u_R^2$. Near the horizon, $u_R=1$, this 
equation is identical to the scalar wave equation in the AdS$_5$ 
Schwarzschild background. In particular, the near-horizon solution
is of the form $\chi\sim (1-u_R)^\alpha$ with $\alpha=\pm i\mathfrak{w}/2$.
The retarded correlation function is related to the solution that 
satisfies infalling boundary conditions, $\alpha=- i\mathfrak{w}/2$,
near the horizon. We have solved equ.~(\ref{scal_wv_eq}) numerically 
by starting from the analytical solution near $u_R=1$, 
\be 
 \chi(\omega,u_R)= \left(1-u_R\right)^{-i\mathfrak{w}/2}\left[ 1 - 
   \frac{\mathfrak{w}}{4(i+\mathfrak{w})}\,
      \left(1-u_R\right) + \ldots \right]\, , 
\ee
and integrating towards the boundary at $u_R=0$. The retarded correlation 
function is given by 
\be 
 G_R(\omega) = \frac{\beta r_+^3\Delta v}{4\pi G_5}\,
  \left. \frac{f(u_R)\chi^\prime(\omega,u_R)}
              {u_R\chi(\omega,u_R)}\right|_{u_R\to 0}\, .  
\ee
where $G_R\equiv (G_R)_{xy,xy}$ and the viscosity spectral function is 
$\eta(\omega)=\frac{1}{\omega}{\it Im} G_R(\omega)$. The spectral function 
is shown in Fig.~\ref{fig_eta_w}. We observe that $\eta(0)/s=1/(4\pi)$ 
\cite{Herzog:2008wg,Adams:2008wt}, and that at large frequency the spectral 
function grows as $\omega^{1/3}$.

\subsection{Low frequency behavior}

 The behavior of the spectral function at large and small $\omega$ can 
be understood analytically. At low frequency we use the ansatz
\be 
\label{w_ans}
\chi(\mathfrak{w},u_R) = 
   (1-u_R)^{-i\mathfrak{w}/2} \left[ 1+ i\mathfrak{w}F(u_R) 
     + \mathfrak{w}^2G(u_R) + \ldots \right]\, . 
\ee
Putting this ansatz into the wave equation determines the functions 
$F$ and $G$. We find
\bea 
 F(u_R) &=&  -\frac{1}{2} \log\left( \frac{1+u_R}{2} \right)\, , \\
 G(u_R) &=&   \frac{1}{8} \left[\log\left( \frac{1+u_R}{2} \right)\right]^2
         - \frac{1}{2} {\it Li}_2\left( \frac{1-u_R}{2} \right)
\eea
Inserting equ.~(\ref{w_ans}) into the boundary action gives the 
retarded correlation function 
\be 
\label{G_R_low}
 G_R(\mathfrak{w}) = -\frac{sT}{2} \left[ i\mathfrak{w} 
   + \mathfrak{w}^2\log(2) + \ldots \right]\, . 
\ee
Matching this result to the Kubo relation (\ref{G_R_Kubo}) leads to 
\be 
\frac{\eta}{s}=\frac{1}{4\pi}\, , \hspace{0.5cm}
\tau_\pi=-\frac{\log(2)}{2\pi T}\, . 
\ee
The sign of the viscous relaxation time is unusual. In ${\cal N}=4$ 
SUSY Yang Mills theory \cite{Baier:2007ix}, as well as in weak coupling 
calculations for dilute gases and the quark gluon plasma the relaxation 
time is always positive \cite{York:2008rr,Schaefer:2014xma}. Equations
(\ref{disp_shear}) and (\ref{disp_sound}) show that $\tau_\pi$ determines 
certain higher order corrections to the dispersion relation of shear
and sound modes, and that in the fluid dynamic regime $\omega\tau_\pi
\ll 1$ there is no constraint on the sign of $\tau_\pi$. On the other 
hand, if we try to match the correlation function to an Israel-Stewart
like scheme, then we find an unstable mode near the ultraviolet 
cutoff.  We note that the $O(\omega^2)$ term in equ.~(\ref{G_R_low}) 
has the same sign as the corresponding term in the ${\cal N}=4$ theory. 
The difference in sign arises from the Kubo relation, which has an extra 
term $G_R \sim -\frac{\kappa_R}{2} \omega^2$ in the relativistic case. 
This term is absent in the non-relativistic theory in both 2+1 and 3+1 
dimensions, and it is also absent in a 2+1 dimensional relativistic theory
\cite{Baier:2007ix}.

\subsection{High frequency behavior}

 The high frequency limit can be studied using a WKB approximation. 
We define $\psi(\mathfrak{w},u_R)=\sqrt{(1-u_R^2)/u_R}\,\chi(\mathfrak{w},
u_R)$. Then  $\psi(\mathfrak{w},u_R)$ satisfies a Schr\"odinger-like equation
\be 
\psi^{\prime\prime}(\mathfrak{w},u_R)
 + \frac{1}{4u_R^2(1-u_R^2)^2} 
  \left( -3 +6u_R^2 + 4\mathfrak{w}^2u_R^3 +u_R ^4\right) 
   \psi (\mathfrak{w},u_R) = 0 \, . 
\ee
For $\mathfrak{w}\gg1$ the function $ \psi (\mathfrak{w},u_R)$ is 
rapidly oscillating in the bulk. We can write 
\be 
\label{psi_wkb}
 \psi (\mathfrak{w},u_R) \simeq \frac{1}{\sqrt{p(u_R)}}
    e^{\pm i(S_0(u_R)+\varphi)} \, , \hspace{0.5cm}
    S_0(u_R) = \int^{u_R} p(u_R')\,du_R'\, . 
\ee
with 
\be 
p(u_R) = \frac{\mathfrak{w}\sqrt{u_R}}{1-u_R^2}\, ,\hspace{0.5cm}
S_0(u_R) = \mathfrak{w} \left[ -{\rm arctan}\left(\sqrt{u_R}\right) 
   +{\rm arctanh}\left(\sqrt{u_R}\right)\right]\, . 
\ee
The two wave functions in equ.~(\ref{psi_wkb}) are linearly independent.
The correct solution is determined by matching to the near horizon 
solution. Near $u_R=1$ we have 
\be 
p(u_R) =\frac{\mathfrak{w}}{2(1-u_R)}\, ,\hspace{0.5cm}
S_0(u_R) = -\frac{\mathfrak{w}}{2} \log(1-u_R)\, . 
\ee
Comparing the WKB result to the infalling solution $\psi(u)\sim 
(1-u_R)^{(1-i\mathfrak{w})/2}$ picks out the ``+'' sign in equ.~(\ref{psi_wkb}). 
Once the sign is fixed the WKB solution determines a unique boundary 
wave function. The general solution near $u_R=0$ is 
\be 
  \psi (\mathfrak{w},u_R) \simeq \frac{1}{\sqrt{u_R}} 
  \left[ c_1 {\rm Ai}'\left((-1)^{1/3}\mathfrak{w}^{2/3} u_R\right) 
       + c_2 {\rm Bi}'\left((-1)^{1/3}\mathfrak{w}^{2/3} u_R\right) 
   \right]\, , 
\ee
where ${\rm Ai}'$ and ${\rm Bi}'$ are derivatives of the Airy 
function. Using the asymptotic behavior of the Airy functions, 
\be
{\rm Ai}'(z) \simeq-\frac{z^{1/4} }{2\sqrt{\pi}}\,e^{-\frac{2}{3}z^{3/2}}\, ,
\hspace{0.5cm}
{\rm Bi}'(z) \simeq \frac{z^{1/4} }{\sqrt{\pi}}\, e^{\frac{2}{3}z^{3/2}}\, ,
\ee
we see that the infalling wave functions matches to the solution proportional
to ${\rm Bi}'$. The corresponding mode function $\chi$, normalized according
to $\chi(\mathfrak{w},0)=1$, is 
\be 
  \chi (\mathfrak{w},u_R) \simeq \frac{\Gamma(1/3)}{3^{1/6}\sqrt{1-u_R^2}} 
  \, {\rm Bi}'\left((-1)^{1/3}\mathfrak{w}^{2/3} u_R\right)\, .  
\ee
Inserting this solution into the boundary action gives
\be 
\frac{\eta(\omega)}{s} \simeq \frac{1}{4\pi}
  \frac{3^{1/6}\Gamma(1/3)}{2\Gamma(2/3)}
 \, \mathfrak{w}^{1/3}\, . 
\ee
This result is shown as the dashed line in Fig.~\ref{fig_eta_w}. 
We observe that the asymptotic behavior sets is rapidly for 
$\mathfrak{w}\gsim 1$. In particular, there are no oscillations 
around the asymptotic form as is the case in the ${\cal N}=4$ 
theory \cite{Teaney:2006nc}.

\begin{figure}[t]
\bc\includegraphics[width=7.5cm]{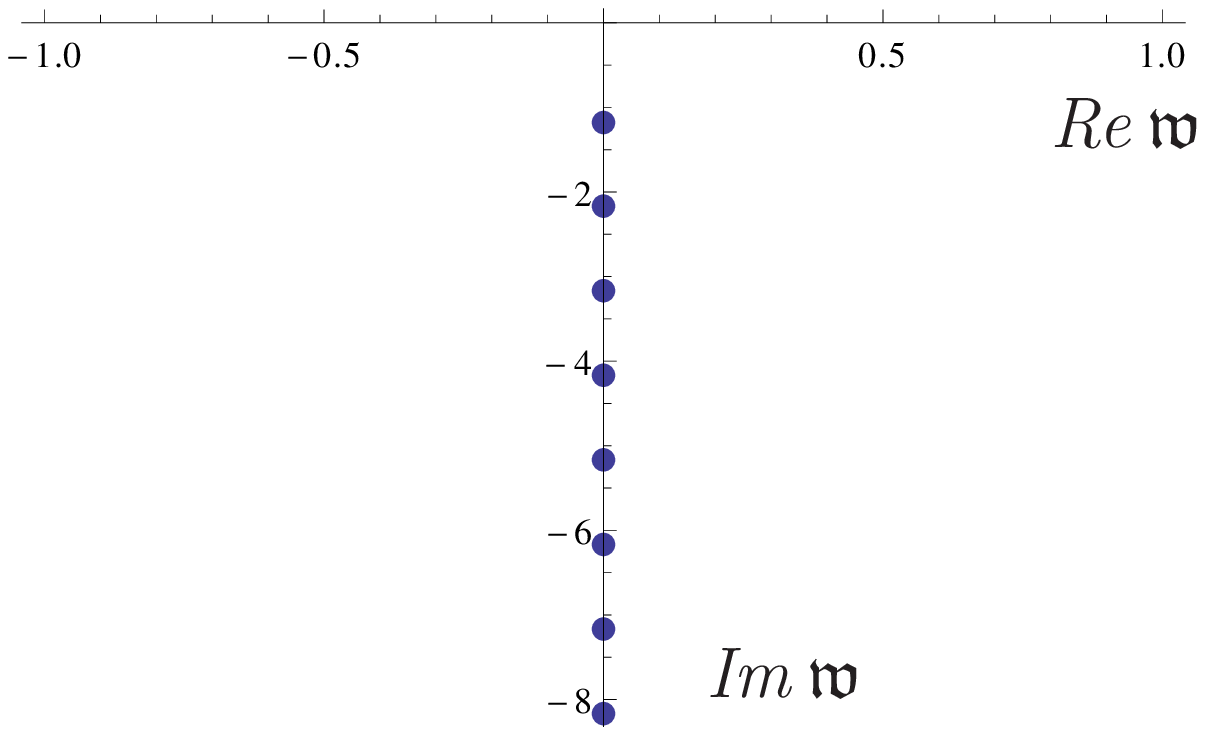}
\includegraphics[width=7.5cm]{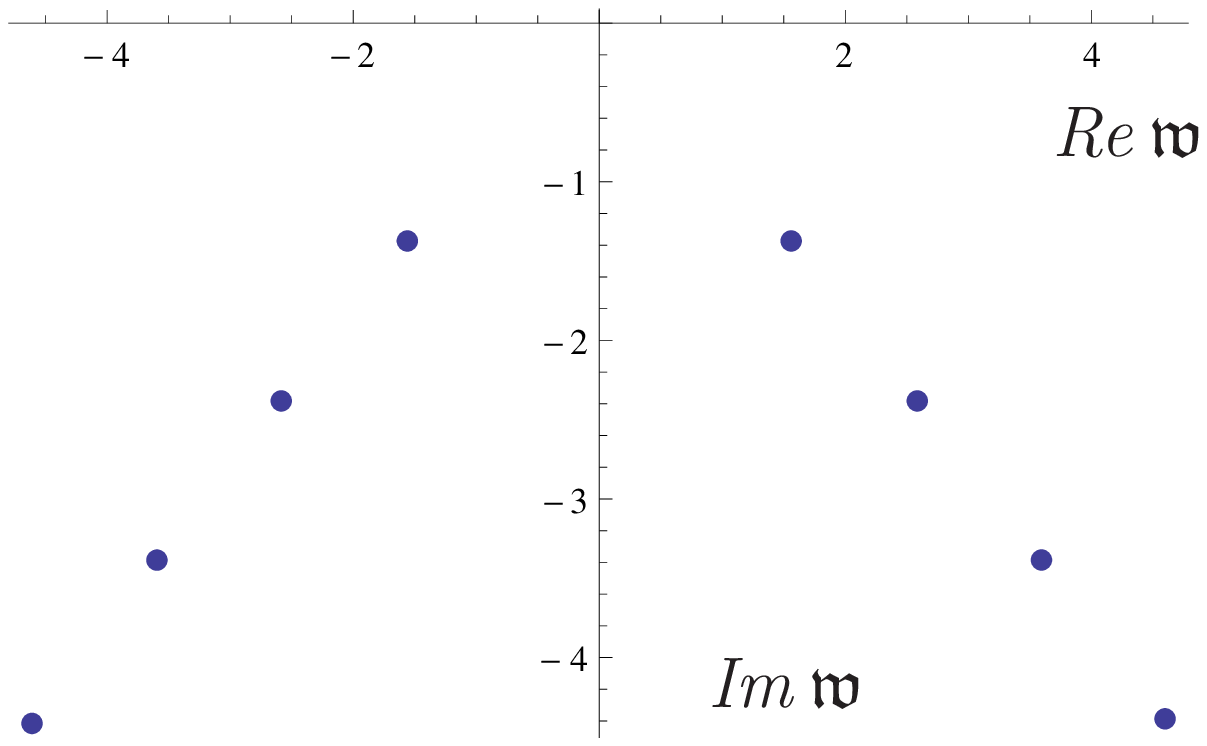}\ec
\caption{\label{fig_qnm}
Complex frequency of quasi-normal modes in the shear channel
for zero spatial momentum. The left panel shows results for
the light-like compactification studied in this work, and the 
right panel shows quasi-normal modes of the  $AdS_5$ Schwarzschild 
black hole \cite{Starinets:2002br}. }   
\end{figure}

\section{Quasi-normal modes}
\label{sec_qnm}

 We have seen that the viscous relaxation time is negative, and 
that in the context of an Israel-Stewart scheme this results implies
the existence of an unstable mode outside the hydrodynamic regime. 
In order to understand relaxation to the hydrodynamic limit, and 
to verify that the fluid is indeed stable, we have computed the
spectrum of quasi-normal modes in the shear channel. We note that 
near $u_R=0$ the solution of the wave equation is of the form 
\be 
 \chi (\mathfrak{w},u_R) \simeq {\cal A}(\mathfrak{w}) 
  \big[ 1+ \ldots \big] + {\cal B}(\mathfrak{w}) 
  \big[ u_R^2 + \ldots \big]\, , 
\ee
and $G_R(\mathfrak{w}) \sim {\cal B}(\mathfrak{w})/{\cal A}(\mathfrak{w})$.
Finding poles of $G_R(\mathfrak{w})$ in the complex $\mathfrak{w}$ plane
corresponds to solutions $\psi (\mathfrak{w},u_R)$ that satisfy a Dirichlet 
problem at $u_R=0$ and infalling boundary conditions at $u_R=1$. We follow
\cite{Starinets:2002br} and solve the Dirichlet problem by transforming 
equ.~(\ref{ads_nr_de}) to the standard form of the Heun differential 
equation. We define 
\be 
 \chi (\mathfrak{w},1-z) = z^{-i\mathfrak{w}/2} (z-2)^{-\mathfrak{w}/2}
   y(z)\, .
\ee
Then $y(z)$ satisfies 
\be 
 y^{\prime\prime}(z)
 + \left[ \frac{\gamma}{z} + \frac{\delta}{z-1} + \frac{\epsilon}{z-2}
   \right] y^{\prime}(z) 
 + \frac{\alpha\beta z-Q}{z(z-1)(z-2)} \, y(z) =0 \, ,
\ee
with $\alpha+\beta=\delta+\gamma+\epsilon-1$ and 
\bea 
&\alpha=\beta=-\frac{\mathfrak{w}}{2}(1+i)\, , \hspace{0.5cm}
 \gamma=1-i\mathfrak{w}\, , \hspace{0.5cm}
 \delta=-1 \, , \hspace{0.5cm}
 \epsilon=1- \mathfrak{w} \, , & \\
& Q = -\frac{\mathfrak{w}}{2}(1-i) +\frac{i}{2}\mathfrak{w}^2\,  . & 
\eea
We seek the solution $y(z)$ as a power series expansion 
\be 
 y(z) = \sum_n a_nz^n \, , 
\ee
where the coefficients $a_n$ satisfy a two-term recursion relation
\be
 a_{n+2} + A_n(\mathfrak{w})a_{n+1} + B_n(\mathfrak{w})a_{n}  = 0\, ,  
\ee
The coefficients $A_n$ and $B_n$ are given by \cite{Starinets:2002br}
\bea 
 A_n &=& - \frac{(n+1)(2\delta+\epsilon+3(n+\gamma))+Q}
                {2(n+2)(n+1+\gamma)}\, ,  \\
 B_n &=& \;\, \frac{(n+\alpha)(n+\beta)}
                {2(n+2)(n+1+\gamma)}\, . 
\eea
The near horizon behavior implies that $a_0=1$ and $a_1=Q/(2\gamma)$.
From the asymptotic behavior of $A_n$ and $B_n$ we see that the series
expansion converges at least for $|z|<1$. We have numerically searched 
for solutions of the Dirichlet problem $y(1)=0$. The results are shown 
in the left panel of Fig.~\ref{fig_qnm}. We observe that the quasi-normal
are located along the negative imaginary axis. This implies, in particular,
that there are no unstable modes. The scale of the first non-hydrodynamic 
mode is set by the temperature, $\omega^1_{\it QNM}\simeq -1.18(2\pi i T)$.
For comparison we also show the quasi-normal modes of the $AdS_5$ black hole 
computed with the same method (right panel). The $AdS_5$ black hole 
quasi-normal modes are governed by the same recursion relation, but with 
a different governing parameter, $Q=-\frac{\mathfrak{w}}{2}(1-i) +\frac{i}{2}
\mathfrak{w}^2(1+2i)$. The difference in $Q$ results in a doubling of the 
quasi-normal mode spectrum, in agreement with the results obtained in 
\cite{Starinets:2002br}. 

\section{Outlook}
\label{sec_out}

 We can compare our results to predictions from kinetic theory and 
quantum many-body theory. The field theory dual of the gravitational
action given in equ.~(\ref{S_5d}) has not been studied, but there 
are a number of general results that do not depend on details of 
the underlying field theory. If the fluid has an effective description
in terms of quasi-particles then the viscosity spectral function can
be studied using kinetic theory. In kinetic theory we find 
\cite{Braby:2010tk,Schafer:2011my}
\be 
 \eta(\omega) = \frac{\eta(0)}{1+\omega^2\tau_R^2}
\ee
where $\tau_R^{-1}$ is the lowest eigenvalue of the linearized 
collision operator. In kinetic theory we also find $\tau_\pi=\tau_R$
\cite{Chao:2011cy,Schaefer:2014xma}, which means that the kinetic
relaxation time is equal to the relaxation time for viscous stresses. 
The viscosity $\eta(0)$ of the dilute Fermi gas was computed in 
\cite{Schafer:2011my,Bruun:2011}, and it was shown that kinetic
theory extrapolated to the BKT transition is consistent with a
shear viscosity to entropy density ratio as small as $1/(4\pi)$.
Hydrodynamic fluctuations lead to a logarithmic divergence of the 
shear viscosity in 2+1 dimensional fluids, $\eta(\omega)\sim 
\log(T/\omega)$ \cite{Chafin:2012eq}. This divergence is suppressed 
in systems with a large number $N$ of internal degrees of freedom. 

  Kinetic theory is applicable for $\omega<T$. The high frequency
tail of the spectral function is determined by the operator product
expansion (OPE) \cite{Hofmann:2011qs,Goldberger:2011hh}. We can 
write 
\be 
\eta(\omega) = \sum_k \langle {\cal O}_k \rangle 
   \, \frac{1}{\omega^{(\Delta_k-d)/2}}\, , 
\ee
where $d$ is the number of spatial dimensions, $\Delta_k$ is the 
dimension of the operator ${\cal O}_k$, and we have set the mass 
$m=1$. In the dilute Fermi gas the leading operator is ${\cal O}_{\cal C}
=\phi^\dagger\phi$ where $\phi=C_0\epsilon_{\alpha\beta}\psi^\alpha
\psi^\beta$ is the difermion operator and $C_0$ is the four-fermion 
coupling. The operator ${\cal O}_{\cal C}$ is known as the contact 
density \cite{Tan:2005}. It has dimension $\Delta_{\cal C}=4$ in 
both $d=3$ and $d=2$ dimensions. This implies that the high 
frequency tail of the spectral function is $\eta(\omega)\sim 1/
\sqrt{\omega}$ in $d=3$, and $\eta (\omega)\sim 1/\omega$ in $d=2$.
Our result $\eta (\omega)\sim \omega^{1/3}$ corresponds to an 
operator of dimension $\Delta=4/3$ in $d=2$.

There are a number of questions that remain to be addressed. The first 
is to understand the asymptotic behavior of the spectral function. 
What is the operator in the dual field theory that governs the power
law? Within a larger class of holographic models, how is the 
asymptotic behavior of the spectral function encoded in the geometry? 
The second set of questions has to do with the unusual sign of the 
viscous relaxation time. How does this sign manifest itself in the 
approach to equilibrium? This can be studied, for example, using the 
fluid-gravity correspondence \cite{Rangamani:2008gi}. Finally, what 
is the significance of the quasi-normal mode spectrum? In particular,
how does the difference between the spectrum for $AdS_5\times S_5$ 
black hole and the Galilean model manifest itself in the relaxation
towards equilibrium? Ultimately, we are also interested in a broader 
class of models that realize Gallilean invariance without using 
lightlike compactifications. Proposals in this direction can be found in 
\cite{Bekaert:2011cu,Janiszewski:2012nf,Janiszewski:2014ewa,Hoyos:2013qna,Eling:2014saa} 

 Acknowledgments: This work was supported in parts by the US Department 
of Energy grant DE-FG02-03ER41260. I would like to thank A.~Karch, 
M.~Kaminski, and P.~Romatschke for useful discussions. This work was
completed at the European Centre for Theoretical Studies in Nuclear 
Physics and Related Areas (ECT*) in Trento (Italy) during the workshop
on Hydrodynamics for Strongly Coupled Fluids, and at the Institute 
for Nuclear Theory (INT) in Seattle.



\begin{thebibliography}{20}

\bibitem{Schafer:2009dj}
T.~Sch\"afer and D.~Teaney,
``Nearly Perfect Fluidity: From Cold Atomic Gases to Hot Quark Gluon
Plasmas,''
Rept.\ Prog.\ Phys.\  {\bf 72}, 126001 (2009)
[arXiv:0904.3107 [hep-ph]].

\bibitem{McGreevy:2009xe} 
J.~McGreevy,
``Holographic duality with a view toward many-body physics,''
Adv.\ High Energy Phys.\  {\bf 2010}, 723105 (2010)
[arXiv:0909.0518 [hep-th]].

\bibitem{Adams:2012th} 
A.~Adams, L.~D.~Carr, T.~Sch\"afer, P.~Steinberg and J.~E.~Thomas,
``Strongly Correlated Quantum Fluids: Ultracold Quantum Gases, 
Quantum Chromodynamic Plasmas, and Holographic Duality,''
New J.\ Phys.\  {\bf 14}, 115009 (2012)
[arXiv:1205.5180 [hep-th]].

\bibitem{DeWolfe:2013cua} 
O.~DeWolfe, S.~S.~Gubser, C.~Rosen and D.~Teaney,
``Heavy ions and string theory,''
Prog.\ Part.\ Nucl.\ Phys.\  {\bf 75}, 86 (2014)
[arXiv:1304.7794 [hep-th]].

\bibitem{Schaefer:2014awa} 
T.~Sch\"afer,
``Fluid Dynamics and Viscosity in Strongly Correlated Fluids,''
Ann.\ Rev.\ Nucl.\ Part.\ Sci., in press (2014)
[arXiv:1403.0653 [hep-ph]].

\bibitem{Son:2008ye} 
D.~T.~Son,
``Toward an AdS/cold atoms correspondence: A Geometric realization of 
the Schrodinger symmetry,''
Phys.\ Rev.\ D {\bf 78}, 046003 (2008)
[arXiv:0804.3972 [hep-th]].

\bibitem{Balasubramanian:2008dm}
K.~Balasubramanian and J.~McGreevy,
``Gravity duals for non-relativistic CFTs,''
Phys.\ Rev.\ Lett.\  {\bf 101}, 061601 (2008)
[arXiv:0804.4053 [hep-th]].

\bibitem{Herzog:2008wg} 
C.~P.~Herzog, M.~Rangamani and S.~F.~Ross,
``Heating up Galilean holography,''
JHEP {\bf 0811}, 080 (2008)
[arXiv:0807.1099 [hep-th]].

\bibitem{Adams:2008wt} 
A.~Adams, K.~Balasubramanian and J.~McGreevy,
``Hot Spacetimes for Cold Atoms,''
JHEP {\bf 0811}, 059 (2008)
[arXiv:0807.1111 [hep-th]].

\bibitem{Maldacena:2008wh}
J.~Maldacena, D.~Martelli and Y.~Tachikawa,
``Comments on string theory backgrounds with non-relativistic conformal
symmetry,''
JHEP {\bf 0810}, 072 (2008)
[arXiv:0807.1100 [hep-th]].

\bibitem{Bloch:2008zzb} 
I.~Bloch, J.~Dalibard and W.~Zwerger,
``Many-body physics with ultracold gases,''
Rev.\ Mod.\ Phys.\  {\bf 80}, 885 (2008)
[arXiv:0704.3011 [cond-mat.other]].

\bibitem{Giorgini:2008zz} 
S.~Giorgini, L.~P.~Pitaevskii and S.~Stringari,
``Theory of ultracold atomic Fermi gases,''
Rev.\ Mod.\ Phys.\  {\bf 80}, 1215 (2008)
[arXiv:0706.3360 [cond-mat.other]].

\bibitem{oHara:2002}
K.~M.~O'Hara, S.~L.~Hemmer, M.~E.~Gehm, S.~R.~Granade, J.~E.~Thomas,
``Observation of a Strongly-Interacting Degenerate Fermi Gas of Atoms,''
Science {\bf 298}, 2179 (2002)
[cond-mat/0212463].

\bibitem{Schafer:2007pr}
T.~Sch\"afer,
``The Shear Viscosity to Entropy Density Ratio of Trapped Fermions in the
Unitarity Limit,''
Phys.\ Rev.\  A {\bf 76}, 063618 (2007)
[arXiv:cond-mat/0701251].

\bibitem{Cao:2010wa}
C.~Cao, E.~Elliott, J.~Joseph, H.~Wu, J.~Petricka, T.~Sch\"afer
and J.~E.~Thomas,
``Universal Quantum Viscosity in a Unitary Fermi Gas,''
Science {331}, 58 (2011)
[arXiv:1007.2625 [cond-mat.quant-gas]].

\bibitem{Elliott:2013b}
E.~Elliott, J.~A.~Joseph, J.~E.~Thomas,
``Anomalous minimum in the shear viscosity of a Fermi gas,''
arXiv:1311.2049 [cond-mat.quant-gas].

\bibitem{Kovtun:2004de}
P.~Kovtun, D.~T.~Son and A.~O.~Starinets,
``Viscosity in strongly interacting quantum field theories from black hole
physics,''
Phys.\ Rev.\ Lett.\  {\bf 94}, 111601 (2005)
[arXiv:hep-th/0405231].

\bibitem{Bruun:2005}
G.~M.~Bruun, H.~Smith,
``Viscosity and thermal relaxation for a resonantly interacting 
Fermi gas,''
Phys.\ Rev.\ A {\bf 72}, 043605 (2005) 
[cond-mat/0504734].

\bibitem{Rupak:2007vp} 
G.~Rupak and T.~Sch\"afer,
``Shear viscosity of a superfluid Fermi gas in the unitarity limit,''
Phys.\ Rev.\ A {\bf 76}, 053607 (2007)
[arXiv:0707.1520 [cond-mat.other]].

\bibitem{Braby:2010tk} 
M.~Braby, J.~Chao and T.~Sch\"afer,
``Viscosity spectral functions of the dilute Fermi gas in kinetic theory,''
New J.\ Phys.\  {\bf 13}, 035014 (2011)
[arXiv:1012.0219 [cond-mat.quant-gas]].

\bibitem{Schaefer:2014xma} 
T.~Sch\"afer,
``Second order fluid dynamics for the unitary Fermi gas from kinetic theory,''
arXiv:1404.6843 [cond-mat.quant-gas].

\bibitem{Taylor:2010ju} 
E.~Taylor and M.~Randeria,
``Viscosity of strongly interacting quantum fluids: spectral functions and 
sum rules,''
Phys.\ Rev.\ A {\bf 81}, 053610 (2010)
[arXiv:1002.0869 [cond-mat.quant-gas]].

\bibitem{Enss:2010qh} 
T.~Enss, R.~Haussmann and W.~Zwerger,
``Viscosity and scale invariance in the unitary Fermi gas,''
Annals Phys.\  {\bf 326}, 770 (2011)
[arXiv:1008.0007 [cond-mat.quant-gas]].
 
\bibitem{Hofmann:2011qs}
J.~Hofmann,
``Current response, structure factor and hydrodynamic quantities 
of a two- and three-dimensional Fermi gas from the operator product 
expansion,''
Phys.\ Rev.\ A {\bf 84}, 043603 (2011)
[arXiv:1106.6035 [cond-mat.quant-gas]].

\bibitem{Goldberger:2011hh} 
W.~D.~Goldberger and Z.~U.~Khandker,
``Viscosity Sum Rules at Large Scattering Lengths,''
Phys.\ Rev.\ A {\bf 85}, 013624 (2012)
[arXiv:1107.1472 [cond-mat.stat-mech]].

\bibitem{Chafin:2012eq} 
C.~Chafin and T.~Sch\"afer,
``Hydrodynamic fluctuations and the minimum shear viscosity of the dilute 
Fermi gas at unitarity,''
Phys.\ Rev.\ A {\bf 87}, 023629 (2013)
[arXiv:1209.1006 [cond-mat.quant-gas]].

\bibitem{Romatschke:2012sf} 
P.~Romatschke and R.~E.~Young,
``Implications of hydrodynamic fluctuations for the minimum shear 
viscosity of the dilute Fermi gas at unitarity,''
Phys.\ Rev.\ A {\bf 87}, no. 5, 053606 (2013)
[arXiv:1209.1604 [cond-mat.quant-gas]].

\bibitem{Wlazlowski:2013owa} 
G.~Wlazlowski, P.~Magierski, A.~Bulgac and K.~J.~Roche,
``The temperature evolution of the shear viscosity in a unitary Fermi gas,''
Phys.\  Rev.\  A 88, {\bf 013639} (2013)
[arXiv:1304.2283 [cond-mat.quant-gas]].

\bibitem{Vogt:2011}
E.~Vogt, M.~Feld, B.~Fr\"ohlich, D.~Pertot, M.~Koschorreck, M.~K\"ohl,
``Scale invariance and viscosity of a two-dimensional Fermi gas,''
Phys.\ Rev.\ Lett.\ {\bf 108}, 070404 (2012)
arXiv:1111.1173 [cond-mat.quant-gas].

\bibitem{Schafer:2011my} 
T.~Sch\"afer,
``Shear viscosity and damping of collective modes in a two-dimensional 
Fermi gas,''
Phys.\ Rev.\ A {\bf 85}, 033623 (2012)
[arXiv:1111.7242 [cond-mat.quant-gas]].

\bibitem{Bruun:2011}
G.~M.~Bruun,
``Shear viscosity and spin diffusion coefficient of a two-dimensional 
Fermi gas,''
Phys.\ Rev.\ A. {\bf 85}, 013636 (2012)
[arXiv:1112.2395 [cond-mat.quant-gas]].

\bibitem{Hofmann:2012np} 
J.~Hofmann,
``Quantum anomaly, universal relations and breathing mode of a 
two-dimensional Fermi gas,''
Phys.\ Rev.\ Lett.\  {\bf 108}, 185303 (2012)
[arXiv:1112.1384 [cond-mat.quant-gas]].

\bibitem{Taylor:2012}
E.~Taylor, M.~Randeria,
``Apparent Low-Energy Scale Invariance in Two-Dimensional Fermi Gases,''
Phys.\ Rev.\ Lett.\ {\bf 109}, 135301 (2012)
[arXiv:1205.1525 [cond-mat.quant-gas]].

\bibitem{Chafin:2013zca} 
C.~Chafin and T.~Sch\"afer,
``Scale breaking and fluid dynamics in a dilute two-dimensional Fermi gas,''
Phys.\ Rev.\ A {\bf 88}, 043636 (2013)
[arXiv:1308.2004 [cond-mat.quant-gas]].

\bibitem{Teaney:2006nc}
D.~Teaney,
``Finite temperature spectral densities of momentum and R-charge  
correlators in $N=4$ Yang Mills theory,''
Phys.\ Rev.\  D {\bf 74}, 045025 (2006)
[arXiv:hep-ph/0602044].

\bibitem{Kovtun:2006pf}
P.~Kovtun and A.~Starinets,
``Thermal spectral functions of strongly coupled N = 4 supersymmetric
Yang-Mills theory,''
Phys.\ Rev.\ Lett.\  {\bf 96}, 131601 (2006)
[arXiv:hep-th/0602059].

\bibitem{Baier:2007ix}
R.~Baier, P.~Romatschke, D.~T.~Son, A.~O.~Starinets and M.~A.~Stephanov,
``Relativistic viscous hydrodynamics, conformal invariance, and holography,''
JHEP {\bf 0804}, 100 (2008)
[arXiv:0712.2451 [hep-th]].

\bibitem{Starinets:2002br} 
A.~O.~Starinets,
``Quasinormal modes of near extremal black branes,''
Phys.\ Rev.\ D {\bf 66}, 124013 (2002)
[hep-th/0207133].

\bibitem{Nunez:2003eq} 
A.~Nunez and A.~O.~Starinets,
``AdS / CFT correspondence, quasinormal modes, and thermal correlators 
in N=4 SYM,''
Phys.\ Rev.\ D {\bf 67}, 124013 (2003)
[hep-th/0302026].

\bibitem{Kovtun:2005ev} 
P.~K.~Kovtun and A.~O.~Starinets,
``Quasinormal modes and holography,''
Phys.\ Rev.\ D {\bf 72}, 086009 (2005)
[hep-th/0506184].

\bibitem{Son:2005tj} 
D.~T.~Son,
``Vanishing bulk viscosities and conformal invariance of unitary Fermi gas,''
Phys.\ Rev.\ Lett.\  {\bf 98}, 020604 (2007)
[cond-mat/0511721].

\bibitem{Chao:2011cy} 
J.~Chao and T.~Sch\"afer,
``Conformal symmetry and non-relativistic second order fluid dynamics,''
Annals Phys.\  {\bf 327}, 1852 (2012)
[arXiv:1108.4979 [hep-th]].

\bibitem{Israel:1979wp}
W.~Israel and J.~M.~Stewart,
``Transient relativistic thermodynamics and kinetic theory,''
Annals Phys.\  {\bf 118}, 341 (1979).

\bibitem{Baker:1999np} 
G.~A.~Baker,
``Neutron matter model,''
Phys.\ Rev.\ C {\bf 60}, 054311 (1999).

\bibitem{York:2008rr}
M.~A.~York and G.~D.~Moore,
``Second order hydrodynamic coefficients from kinetic theory,''
Phys.\ Rev.\ D {\bf 79}, 054011 (2009)
[arXiv:0811.0729 [hep-ph]].

\bibitem{Tan:2005}
S.~Tan, 
``Large momentum part of fermions with large scattering length,''
Ann.\ Phys.\ {\bf 323}, 2971 (2008).

\bibitem{Rangamani:2008gi} 
M.~Rangamani, S.~F.~Ross, D.~T.~Son and E.~G.~Thompson,
``Conformal non-relativistic hydrodynamics from gravity,''
JHEP {\bf 0901}, 075 (2009)
[arXiv:0811.2049 [hep-th]].

\bibitem{Bekaert:2011cu} 
X.~Bekaert, E.~Meunier and S.~Moroz,
``Towards a gravity dual of the unitary Fermi gas,''
Phys.\ Rev.\ D {\bf 85}, 106001 (2012)
[arXiv:1111.1082 [hep-th]].

\bibitem{Janiszewski:2012nf} 
S.~Janiszewski and A.~Karch,
``String Theory Embeddings of Nonrelativistic Field Theories and Their 
Holographic Horava Gravity Duals,''
Phys.\ Rev.\ Lett.\  {\bf 110}, 081601 (2013)
[arXiv:1211.0010 [hep-th]].

\bibitem{Janiszewski:2014ewa} 
S.~Janiszewski, A.~Karch, B.~Robinson and D.~Sommer,
``Charged black holes in Horava gravity,''
arXiv:1401.6479 [hep-th].

\bibitem{Hoyos:2013qna} 
C.~Hoyos, B.~S.~Kim and Y.~Oz,
``Lifshitz Field Theories at Non-Zero Temperature, Hydrodynamics and Gravity,''
JHEP {\bf 1403}, 029 (2014)
[arXiv:1309.6794 [hep-th], arXiv:1309.6794].

\bibitem{Eling:2014saa} 
C.~Eling and Y.~Oz,
``Horava-Lifshitz Black Hole Hydrodynamics,''
arXiv:1408.0268 [hep-th].

\end{thebibliography}
\end{document}